\documentclass[prl,twocolumn,aps,superscriptaddress]{revtex4-1}

\begin{document}

\title{Comment on ``Exact bosonization for an interacting Fermi gas in arbitrary dimensions''}

\author{D. Galanakis}
\author{S. Yang}
\affiliation{Department of Physics and Astronomy, Baton Rouge, LA, USA}

\author{F.F. Assaad}
\affiliation{Institut f\"ur theoretische Physik und Astrophysik, Universit\"at W\"urzburg, Am Hubland, D-97074, Germany}

\author{M. Jarrell}
\affiliation{Department of Physics and Astronomy, Baton Rouge, LA, USA}

\author{P. Werner}
\author{M. Troyer}

\affiliation{Theoretische Physik, ETH Zurich, 8093, Zurich, Switzerland}

\maketitle
In a recent Letter \cite{efetov} Efetov {\it et al.} propose an exact
mapping of an interacting fermion system onto a new model that is supposed to allow
sign-problem free Monte Carlo simulations. In this Comment, we show that their formalism
is equivalent to the standard approach of Blackenbecker, Scalapino and Sugar (BSS) \cite{bss} for fermionic systems and has the same sign statistics and minus sign problem. 

Our first observation is that the partition function for a given configuration of the auxiliary fields $\phi$ is the same in the standard formulation $Z_f$ [Eq. (8) in Ref.~\cite{efetov}] and in their new bosonized scheme $Z_b$ [Eq. (9)]: 
\begin{equation}
Z_{f}\left[\phi\right]=Z_{b}\left[\phi\right].
\end{equation}
This observation is trivial in the limit of the time step $\Delta\tau\rightarrow0$, where both schemes reproduce the same partition function. Since $Z_f$ can be negative also in this limit \cite{bss}, $Z_b$ is also not sign-positive. Both $Z_f$ and $Z_b$ are positive if $\phi$ is a smooth path \cite{bss}, but restricting the configuration space to smooth paths amounts to a semi-classical approximation.

We next show that $Z_{f}\left[\phi\right]=Z_{b}\left[\phi\right]$ also holds for finite $\Delta\tau$ and piecewise constant paths where the field $\phi_{r}(\tau)$
is constant on the interval $\left[(l+1)\Delta,l\Delta\right]$.
Efetov {\it et al.}'s Eq. (9) is equivalent to $Z_{b}\left[\phi\right]={\rm Tr}\left(e^{-\beta H_{0}}U_{I}(\beta,0)\right)$
where $\frac{\partial}{\partial\tau_{1}}U_{I}(\tau_{1},\tau_{0})=-H_{I,1}(\tau_{1})U_{I}(\tau_{1},\tau_{0})$
and $H_{I,1}(\tau)=-\sum_{r}\phi_{r}(\tau)e^{\tau H_{0}}m_{z,r}e^{-\tau H_{0}}$.
Since $U_{I}(\tau_{2},\tau_{0})=U_{I}(\tau_{2},\tau_{1})U_{I}(\tau_{1},\tau_{0})$,
$Z_{b}\left[\phi\right]={\rm Tr}\left[e^{-\beta H_{0}}\prod_{n=0}^{N}U_{I}(l\Delta,(l-1)\Delta)\right]$.
We are left with the task of evaluating $U_{I}\left(l\Delta,(l-1)\Delta\right)$
on the $l^{th}$ time interval where the fields are constant and take
the values $\phi_{rl}$. For this constant in time field configuration,
$U_{I}(l\Delta,(l-1)\Delta)=e^{l\Delta H_{0}}e^{-\Delta h_{l}}e^{-(l-1)\Delta H_{0}}$
with $H_{l}=H_{0}-\sum_{r}\phi_{r,l}m_{z,r}$. Hence, \begin{equation}
Z_{b}\left[\phi\right]=\det\left[1+\prod_{l=0}^{N}e^{-\Delta h_{l}}\right]\equiv Z_{f}\left[\phi\right].\end{equation}

We also demonstrate that the proposed formalism is equivalent to that of BSS \cite{bss}. Starting from
the expression for $Z$ found between Eq. (9) and Eq. (10)  and using a 
matrix notation the partition function reads\begin{equation}
\ln\frac{Z_{b}\left[\phi\right]}{Z_{0}}=\int_{0}^{1}du\sum_{\sigma}Tr\Phi G_{\sigma},\label{eq:partition_function_trPhiG}\end{equation}
where $\Phi$ is a matrix with elements $\Phi\left(rl,r^{\prime}l^{\prime}\right)=\phi_{r,l}\delta_{rr^{\prime}}\delta_{ll^{\prime}}$,
the trace is over the space time indices and \begin{equation}
G_{\sigma}=G^{0}+u\sigma G^{0}\Phi G_{\sigma}=\left(1-u\sigma G^{0}\Phi\right)^{-1}G^{0},\label{eq:Green_function_from_G0}\end{equation}
where $G^{0}$ is the non interacting Green function. From Eqs. (\ref{eq:partition_function_trPhiG})
and (\ref{eq:Green_function_from_G0}) we get
\begin{eqnarray}
\ln\frac{Z_{b}\left[\phi\right]}{Z_{0}} & = & \int_{0}^{1}du\sum_{\sigma}Tr\frac{1}{1-u\sigma G^{0}\Phi}\sigma G^{0}\Phi.\label{eq:bss_partition_1}
 \end{eqnarray}
 The integral in terms of $u$ and $\sigma$ can be carried out analytically
yielding the BSS partition function
\begin{eqnarray}
Z_{b}\left[\phi\right] & = & Z_{0}\exp\left(-Tr\left[\ln\left(1-(G^{0}\Phi)^{2}\right)\right]\right)\nonumber \\
 & = & \frac{Z_{0}}{\det\left(1-(G^{0}\Phi)^{2}\right)}=\prod_{\sigma=\uparrow\downarrow}\frac{1}{\det G_{\sigma}}.\label{eq:partition_function_is_bss}
 \end{eqnarray}
 
A sign problem is present in Eq. (\ref{eq:partition_function_trPhiG}),
but is hidden in the $u$ integral. If $G^{0}\Phi$ has eigenvalues
on the real axis with absolute value greater than 1, then the $u$
integral runs over poles. This will give phases of $i\pi$ in the
exponent which can lead to negative values of the exponential. 

Finally,  Efetov's  {\it et al.} suggest an alternative way of evaluating the same weight $Z_b$ from a 
bosonic field $A_{rr^{\prime}}\left(\tau\right)$;  however,
the equation of motion that $A$ satisfies, [their Eq. (13)], is
singular. Furthermore, the equation of motion 
cannot uniquely determine $A$ -- not even with their additional constraint $\sum_{r}A_{rr}\left(\tau\right)=0$. The latter can be shown, for example, using a two site cluster and a single time slice.

In conclusion, we have shown that for piecewise continuous paths the partition
function obtained from Efetov {\it et al.}'s method \cite{efetov} is equivalent
to the standard BSS formulation and has, in particular, the same sign
problem. However, their method provides a new perspective on the
minus sign problem, as it can be viewed as originating from poles
of the coupling constant ($u$) integral, or as branch cuts of a logarithm
when the integral over $u$ is performed analytically.

\end{document}